\documentclass[11pt]{article}
\usepackage{latexsym}
\usepackage{colortbl}
\usepackage{amssymb}
\usepackage{amsmath}

\usepackage{latexsym}
\usepackage{plain}

\usepackage[german,english]{babel}
\usepackage[T1]{fontenc}
\usepackage[latin1]{inputenc}

\usepackage{amsfonts}
\usepackage{amsmath}
\usepackage{mathrsfs}
\usepackage{psfrag}
\usepackage{graphicx}
\usepackage{epstopdf}
	\DeclareGraphicsExtensions{.eps}
\usepackage{multicol}
\usepackage{caption}
\usepackage{amssymb}
\usepackage{array}
\usepackage{rotating}

\newtheorem{thm}{Theorem}[section]

\numberwithin{equation}{section}

\begin{document}

\vspace*{.5cm}
\begin{center}


{\LARGE{\textbf{An extension of the Ky Fan inequality}}}
\bigskip

{\Large{\bf{Yuri Suhov\footnote{Math Dept, Penn State University,
PA 16802, USA; DPMMS, University of Cambridge, CB30WB, UK; IITP,
RAS, 127994 Moscow GSP-4, Russia. E-mail: ims14@ps.edu, yms@statslab.cam.ac.uk},\;\; Salimeh Yasaei Sekeh\footnote{Department\;of\;Statistics,\;Federal\;University\;of\;S$\tilde{\rm a}$o\;Carlos\;(UFSCar),\;S$\tilde{\rm a}$o\;Carlos,\;Brazil.\;E-mail: sa$_{-}$yasaei@yahoo.com}}}}


\end{center}
\date{\today}
\bigskip
\begin{abstract}
The aim of this paper is to analyze the weighted KyFan inequality proposed in \cite{SY}. A number of numerical simulations involving the exponential weighted function is given. We show that in several cases and types of examples one can imply an improvement of the standard KyFan inequality.
\end{abstract}
\vskip .5 truecm


\vskip .5 truecm


\vspace{0.5cm}

\mbox{\quad}

\def\fB{\mathfrak B}\def\fM{\mathfrak M}\def\fX{\mathfrak X}
 \def\cB{\mathcal B}  \def\cC{\mathcal C} \def\cM{\mathcal M}\def\cZ{\mathcal Z}
\def\bu{\mathbf u}\def\bv{\mathbf v}\def\bx{\mathbf x}\def\by{\mathbf y}
\def\om{\omega} \def\Om{\Omega}
\def\bbP{\mathbb P} \def\hw{h^{\rm w}} \def\hwphi{{h^{\rm w}_\phi}}
\def\beq{\begin{eqnarray}} \def\eeq{\end{eqnarray}}
\def\beqq{\begin{eqnarray*}} \def\eeqq{\end{eqnarray*}}
\def\rd{{\rm d}} \def\Dwphi{{D^{\rm w}_\phi}}
\def\Lam{\Lambda}
\def\mwe{{D^{\rm w}_\phi}}
\def\DwPhi{{D^{\rm w}_\Phi}} \def\iw{i^{\rm w}_{\phi}}
\def\bE{\mathbb{E}}
\def\1{{\mathbf 1}} \def\fB{{\mathfrak B}}  \def\fM{{\mathfrak M}}
\def\diy{\displaystyle} \def\bbE{{\mathbb E}} \def\bu{\mathbf u}


\def\BA{{\mathbf A}} \def\BC{{\mathbf C}} \def\BI{{\mathbf I}}
\def\BU{{\mathbf U}} \def\BV{{\mathbf V}}
\def\BX{\mathbf{X}} \def\uX{{\underline{\BX_{}}}} \def\BY{{\mathbf Y}}
\def\cY{\mathcal Y}  \def\ux{{\underline{\bx_{}}}}
\def\bPsi{\mathbf{\Psi}}
\def\bsig{mathbf{\sigma}}\def\ov{\overline} \def\oS{{S^{\rm c}}}

\def\fB{\mathfrak B}\def\fM{\mathfrak M}\def\fX{\mathfrak X} \def\fY{\mathfrak Y}
\def\fZ{\mathfrak Z}
 \def\cB{\mathcal B}\def\cM{\mathcal M}\def\cX{\mathcal X} \def\cY{\mathcal Y}
\def\bu{\mathbf u}\def\bv{\mathbf v}\def\bx{\mathbf x}  \def\by{\mathbf y} \def\bz{\mathbf z}
\def\om{\omega} \def\Om{\Omega}
\def\bbP{\mathbb P} \def\bbR{\mathbb R} \def\hw{h^{\rm w}} \def\hwphi{{h^{\rm w}_\phi}}
\def\beq{\begin{eqnarray}} \def\eeq{\end{eqnarray}}
\def\beqq{\begin{eqnarray*}} \def\eeqq{\end{eqnarray*}}
\def\rd{{\rm d}} \def\Dwphi{{D^{\rm w}_\phi}}
\def\Lam{\Lambda} \def\Ups{\Upsilon}
\def\mwe{{D^{\rm w}_\phi}}
\def\DwPhi{{D^{\rm w}_\Phi}} \def\iw{i^{\rm w}_{\phi}}
\def\bE{\mathbb{E}}
\def\1{{\mathbf 1}} \def\fB{{\mathfrak B}}  \def\fM{{\mathfrak M}}
\def\diy{\displaystyle} \def\bbE{{\mathbb E}} \def\bu{\mathbf u}
\def\lam{\lambda} \def\bbB{{\mathbb B}}
\def\bbR{{\mathbb R}}\def\bbS{{\mathbb S}} \def\bmu{{\mbox{\boldmath${\mu}$}}}
 \def\bPhi{{\mbox{\boldmath${\Phi}$}}}
 \def\bbZ{{\mathbb Z}} \def\fF{\mathfrak F}\def\bt{\mathbf t}\def\B1{\mathbf 1}
 \def\b0{\mathbf 0} \def\beacl{\begin{array}{cl}}  \def\beal{\begin{array}{l}}
 \def\beac{\begin{array}{c}}  \def\ena{\end{array}}

\def\uX{{\underline{X_{}}}}  \def\ux{{\underline{x_{}}}}

\def\BA{{\mathbf A}} \def\BB{{\mathbf B}} \def\BC{{\mathbf C}}
\def\BD{{\mathbf D}} \def\BI{{\mathbf I}} \def\BK{{\mathbf K}}
\def\BU{{\mathbf U}} \def\BV{{\mathbf V}}
\def\BX{\mathbf{X}} \def\uX{{\underline{X_{}}}} \def\BY{{\mathbf Y}}
\def\cY{\mathcal Y}  \def\ux{{\underline{x_{}}}}
\def\bPsi{\mathbf{\Psi}}
\def\bsig{mathbf{\sigma}}\def\ov{\overline} \def\oS{{S^{\rm c}}}
\def\beal{\begin{array}{l}} \def\ena{\end{array}}
\def\BG{\mathbf{G}} \def\BE{\mathbf{E}}\def\Bsigma{\mathbf{\sigma}}
\def\wt{\widetilde} \def\rA{{\rm A}} \def\rT{{\rm T}}

\section{Introduction. The weighted Ky Fan inequality}

{\bf 1.1.} The well-known Ky Fan inequality \cite{KF1, KF2, KF3} asserts that $\log\,{\rm{det}}\,\BC$ is a
concave function of a (strictly)
positive definite matrix $\BC$. In other words, $\forall$ strictly positive-definite
$d\times d$ matrices $\BC_1$, $\BC_2$ and $\lam_1,\lam_2\geq 0$ with $\lam_1+\lam_2=1$,
\beq\label{eq:KF} \log{\rm{det}}\Big(\lam_1\BC_1+\lam_2\BC_2\Big)-\sum\limits_{a=1,2}
\lam_a\log{\rm{det}}\,\BC_a\geq 0;\;\hbox{equality iff
$\lam_1\lam_2=0$ or $\BC_1=\BC_2$.}\eeq
For original `geometric` proofs of \eqref{eq:KF} and other related inequalities, see Ref \cite{Mo} and the
bibliography therein.  In \cite{CT, DCT, CTb} the derivation of \eqref{eq:KF} occupies few lines and is
based on properties of information-theoretical entropies; a similar method allows to derive a number of
other determinant-related inequalities.

More precisely, \eqref{eq:KF} is equivalent to the bound for Shannon differential entropies for
Gaussian probability density functions (PDFs):
\beq\label{eq:sKFi} h(f^{\rm{No}}_{\lam_1\BC_1+\lam_2\BC_2})- \lam_1 h(f^{\rm{No}}_{\BC_1})
-\lam_2 h(f^{\rm{No}}_{\BC_2})\geq 0;\hbox{ equality iff $\lam_1\lam_2=0$ of $\BC_1=\BC_2$.}\eeq
Here and below, $f^{\rm{No}}_{\BC}$ stands for the $d$-variate normal PDF ${\rm N}(\b0,\BC)$, with
mean $\b0$ and covariance matrix $\BC$:
\beq\label{eq:fBCi} f^{\rm{No}}_\BC (\bx_1^d):=\frac{1}{(2\pi )^{d/2}\big({\rm{det}\,\BC\big)^{1/2}}}\exp\,\left(
-\frac{1}{2}\,{\bx_1^d}^{\rm T}\BC^{-1}\bx_1^d\right),\;\;\bx_1^d=\begin{pmatrix}x_1\\ \vdots\\ x_n
\end{pmatrix} \in\bbR^d.\eeq
Next, $h(f)=-\diy\int_{\bbR^d}f(\bx_1^d)\log\,f(\bx_1^d)\rd\bx_1^d$ represents the Shannon differential
entropy of a PDF $f$.  In the Gaussian case,
\beq\label{eq:sde} h(f^{\rm{No}}_\BC):=-\int_{\bbR^d}f^{\rm{No}}_\BC(\bx_1^d\log\,f^{\rm{No}}\BC
(\bx_1^d)\rd\bx_1^d =\frac{1}{2}\log \left[ (2\pi )^d ({\rm{det}}\,\BC)\right]
+\frac{d\log\,e}{2}.\eeq
Inequality \eqref{eq:sKFi} is a consequence of the fact that, under certain conditions, $h(f)$ is maximized
$f=f^{\rm{No}}_\BC$. Throughout the paper, we use the abbreviation KFI for either of  \eqref{eq:KF},
\eqref{eq:sKFi}. (Sometimes the term a standard KFI is also used.)
\vskip 1 truecm

{\bf 1.2.} In this paper we compare inequalities \eqref{eq:KF}, \eqref{eq:sKFi} with {\it weighted} inequalities
similar to \eqref{eq:sKFi} and established for {\it weighted} differential entropies in the recent paper
\cite{SY}; see below. For the sake
of pre-emptiveness, we call each of these inequalities a {\it weighted} Ky Fan inequality (WKFI, for short).
A WKFI is obtained for a given non-negative weight function; when this function equals $1$,
the WKFI coincides with the KFI. A natural question is whether a WKFI can provide an `improvement` to
KFI; for instance, by producing a positive lower bound for the LHS in \eqref{eq:KF}, \eqref{eq:sKFi}. We
give a numerical evidence that the answer can be yes or no, depending on the choice of $\BC_a$ and
$\lam_a$. We work with so-called exponential weight functions for which all calculations simplify.
Furthermore, the numerical simulations are done for $d=2$, allowing a graphical representation of
results.

Let $\bx_1^d\in\bbR^d\mapsto\phi (\bx_1^d)\geq 0$ be a given non-negative measurable function positive
on an open domain  in $\bbR^d$. Following \cite{BG}, \cite{C} and \cite{SY}, under the usual agreement,
$0\cdot\log\,0=0\cdot\log\,(+\infty )=0$, the {\it weighted} differential entropy (WDE) of  PDF $f$
with {\it weight function} (WF) $\phi$ is defined by
\beq\label{eq:wde} \hw_\phi (f):=-\int_{\bbR^d}\phi (\bx_1^d)f(\bx_1^d)
\log\,f (\bx_1^d)\rd\bx_1^d,\eeq
assuming that the integral is absolutely convergent. Cf. \cite{BG, C, G}.

In the Gaussian case, the WDE $\hw_\phi (f^{\rm{No}}_\BC)$ admits a representation extending the
RHS in \eqref{eq:sde}. Define a number $\alpha (\BC)=\alpha_\phi (\BC)>0$ and a $d\times d$ matrix
$\bPhi_\BC =\bPhi^{\rm{No}}_{\phi, \BC}$ involving WF $\phi$ and PDF $f^{\rm{No}}_\BC$:
\beq\label{eq:bPhi}\alpha (\BC)=\int\limits_{\bbR^d}\phi (\bx_1^d)
f^{\rm{No}}_{\BC}(\bx_1^d)\rd\bx_1^d,\;\; \bPhi^{\rm{No}}_\BC=\int\limits_{\bbR^d}\bx_1^d\,\left(\bx_1^d
\right)^{\rm T}\phi (\bx_1^d)
f^{\rm{No}}_{\BC}(\bx_1^d)\rd\bx_1^d.\eeq
Then 
\beq\label{eq:sigm}\hwphi (f^{\rm{No}}_{\BC})=\frac{\alpha (\BC)}{2} \log \left[ (2\pi)^d ({\rm{det}}\,\BC)\right]
+\frac{\log\,e}{2}{\rm{tr}}\,\BC^{-1}\bPhi_\BC:=\sigma_\phi (\BC).\eeq
\vskip 1 truecm

{\bf 1.3.} The following theorem was proven in \cite{SY}.
\vskip .5 truecm

\begin{thm}\label{KyFan}
{\rm (The WKFI; cf. \cite{SY}, Theorem 3.2).}  Given
$\lam_1,\lam_2\in [0,1]$ with $\lam_1+\lam_2=1$ and (strictly) positive-definite $\BC_1$, $\BC_2$, set:
 $\BC=\lam_1\BC_1+\lam_2\BC_2$ and $\bPsi
=\lam_1\bPhi_{\BC_1}+\lam_2\bPhi_{\BC_2}-\bPhi_\BC$.
Assume that, ofr a given WF $\phi$,
\beq\label{eq:KyFanCond}\begin{array}{c}
\diy\lam_1\alpha (\BC_1)+\lam_2\alpha (\BC_2)-\alpha (\BC )\geq 0,\\
\diy\Big[\lam_1\alpha (\BC_1)+\lam_2\alpha (\BC_2)-\alpha (\BC )\Big]
\log \left[ (2\pi)^d ({\rm{det}}\,\BC)\right]
 +{\rm{tr}}\,\Big(\BC^{-1}\bPsi\Big)\leq 0.\end{array}
\eeq
Then
\beq\label{eq:wKF}\sigma_\phi (\lam_1\BC_1+\lam_2\BC_2)-\lam_1\sigma_\phi (\BC_1)
-\lam_2\sigma_\phi (\BC_2)\geq 0;\;\hbox{ equality iff $\;\lam_1\lam_2=0\;$ or $\;\BC_1=\BC_2$.}
\eeq
\end{thm}
\vskip .5 truecm

Observe that when $\phi (x)\equiv 1$, bounds \eqref{eq:KyFanCond} are fulfilled for all choices
of $\BC_a$ and $\lam_a$, $a=1,2$. (In fact, they become equalities.) In this case, inequality \eqref{eq:wKF}
coincides with \eqref{eq:sKFi}.

\section{Exponential weight functions}

\def\kap{\kappa} \def\ukap{{\underline\kap}}\def\beac{\begin{array}{c}}

{\bf 2.1.} As was said, in this paper we deal with {\it exponential} WFs, of the form
\beq\label{eq:expophi}\phi (\bx_1^d)=\exp\,\Big({\bt_1^d}^{\rm T}
\bx_1^d\Big)\;\hbox{ where $\bt_1^d\in\bbR^d$.}\eeq
To shorten the notation, we write from now on $\bx$ and $\bt$
instead of $\bx_1^d$ and $\bt_1^d$. Here we use the Laplace transform formulas: for
$\phi (\bx)=\exp\,\Big({\bt}^{\rm T}\bx\Big)$ the Eqn \eqref{eq:bPhi} yields
\beq\label{eq:Lapl1}
\alpha (\BC )=\exp\,\Big(\frac{1}{2}{\bt}^{\rm T}\BC\bt \Big)\;\hbox{ and }\;
\Phi_\BC=\BC\exp\,\Big(\frac{1}{2}{\bt}^{\rm T}\BC\bt \Big).\eeq
Hence, for $\phi (\bx)=\exp\,\Big({\bt}^{\rm T}\bx\Big)$, the WDE \eqref{eq:sigm} becomes
\beq\label{eq:hwphiL} \sigma_{\bt}(\BC)=h(f^{\rm{No}}_\BC)\exp\,\Big(\frac{1}{2}{\bt}^{\rm T}\BC\bt \Big)\eeq
where $h(f^{\rm{No}}_\BC)$ is as in Eqn \eqref{eq:sde}. 

To tackle condition \eqref{eq:KyFanCond}, we introduce the set  $\bbS =\bbS
(\BC_1,\BC_2;\lam_1,\lam_2)\subset\bbR^d$:
\beq\label{eq:KyFanCond0}\begin{array}{l}\diy
\bbS =\Big\{\bt\in\bbR^d:\;F^{(1)}(\bt )\geq 0,\;\hbox{\sl{and}}\;Fö^{(2)}(\bt )\leq 0\Big\}
\;\;\hbox{\sl{where}}\\
\quad\diy F^{(1)}(\bt )=\sum\limits_{\alpha =1,2}\lam_\alpha\exp\,\Big(\frac{1}{2}{\bt}^{\rm T}
\BC_\alpha\bt \Big)
-\exp\,\Big(\frac{1}{2}{\bt}^{\rm T}\BC\bt \Big)\;\hbox{\sl{and}}\\
\quad\diy F^{(2)}(\bt )=\Bigg[\sum\limits_{\alpha =1,2}\lam_\alpha\exp\,\Big(\frac{1}{2}{\bt}^{\rm T}
\BC_\alpha\bt \Big)-\exp\,\Big(\frac{1}{2}{\bt}^{\rm T}\BC\bt \Big)\Bigg]\log \left[ (2\pi)^d ({\rm{det}}\,\BC)
\right]\\
\qquad\qquad\quad +\diy\sum\limits_{\alpha =1,2}\lam_\alpha\exp\,\Big(\frac{1}{2}{\bt}^{\rm T}\BC_\alpha
\bt \Big){\rm{tr}}\,\big[\BC^{-1}\BC_\alpha\big]-d\exp\,\Big(\frac{1}{2}{\bt}^{\rm T}\BC\bt \Big).\end{array}
\eeq

Theorem \ref{KyFan} is transformed into Theorem \ref{KF.Ex.WF}:
\vskip .5 truecm

\begin{thm}\label{KF.Ex.WF}
Given positive definite matrices $\BC_1$, $\BC_2$ and $\lam_1,\lam_2\in [0,1]$
with $\lam_1+\lam_2=1$, set $\BC=\lam_1\BC_1+\lam_2\BC_2$. Assume that $\bt\in\bbS
$. Then
\beq\label{eq:KyFan0}h(f^{\rm{No}}_\BC)\exp\,\Big(\frac{1}{2}{\bt}^{\rm T}\BC\bt \Big)-\lam_1
h(f^{\rm{No}}_{\BC_1})\exp\,\Big(\frac{1}{2}{\bt}^{\rm T}\BC_1\bt \Big)
-\lam_2h(f^{\rm{No}}_{\BC_1})\exp\,\Big(\frac{1}{2}{\bt}^{\rm T}\BC_2\bt \Big)\geq 0;
\eeq
equality iff $\;\lam_1\lam_2=0\;$ or $\;\BC_1=\BC_2$.
\end{thm}
\vskip .5 truecm

{\bf 2.2.} Consequently, $\forall$ $\bt\in\bbS$ we have that
\beq\label{eq:wKFL0}\begin{array}{l}\diy\Sigma (\bt):=\left\{\log \left[ (2\pi e)^d ({\rm{det}}\,\BC)
\right]\right\}
\exp\,\Big(\frac{1}{2}{\bt}^{\rm T}\BC\bt \Big)\\
\qquad\diy
 -\sum\limits_{a =1,2}\lam_a
\left\{\log \left[ (2\pi e)^d ({\rm{det}}\,\BC_a)\right]\right\}
\exp\,\Big(\frac{1}{2}{\bt}^{\rm T}\BC_a\bt \Big)\geq 0.\end{array}\eeq

Eqn \eqref{eq:wKFL0} can be called an exponentially weighted (or briefly: an exponential) Ky Fan inequality (EKFI). When $\bt =0$, the EKFI
is reduced to the standard KFI \eqref{eq:KF}, \eqref{eq:sKFi}. In particular, for $\bt =0$, the bounds
in \eqref{eq:KyFanCond0} become equalities
for any $\BC_a$ and $\lam_a$. (Hence, point $\bt ={\mathbf 0}$
lies in $\bbS$ for any choice of $\BC_a$ and $\lam_a$.)
Consequently, it makes sense to analyze the difference $\Lam (\bt):=\Sigma (\bt)-\Sigma ({\mathbf 0})$:
\beq\label{eq:wKFL2}\begin{array}{l}
\diy\Lam(\bt )(=\Lam (\bt ;\BC_1,\BC_2;\lam_1,\lam_2))= \log\,({\rm{det}}\,\BC)\left[\exp\,\Big(\frac{1}{2}{\bt}^{\rm T}\BC\bt \Big)-1\right]\\
\qquad\qquad\qquad\diy +\sum\limits_{\alpha =1,2}\lam_\alpha
\log\, ({\rm{det}}\,\BC_\alpha)\left[1-\exp\,\Big(\frac{1}{2}{\bt}^{\rm T}\BC_\alpha\bt \Big)\right]\\
\qquad\qquad +\diy d\log(2\pi e)\left[-\sum\limits_{\alpha =1,2}\lam_\alpha
\exp\,\Big(\frac{1}{2}{\bt}^{\rm T}\BC_\alpha\bt \Big) +\exp\,\Big(\frac{1}{2}{\bt}^{\rm T}\BC\bt \Big)\right].\end{array}\eeq
If $\Lam (\bt)>0$, we can think of an improvement in the standard KFI, and if  $\Lam (\bt)<0$,
of a deterioration.



{\bf 2.3.} It has to be said that set $\;\bbS\;$ looks rather involved, and there is no guaranty that it
is not empty.
(In fact, numerical evidence suggests that $\bbS=\emptyset$ for some choices of $\BC_a$ and
$\lam_a$.) Therefore, it makes sense to explore the behavior of $\Lam (\bt )$ for $\bt$ in the whole
of $\bbR^d$. In particular, a stationary point $\bt$ satisfying ${\rm{grad}}_{\bt}\Lam =0$ is found from
\beq\label{eq:statpnt}\beal\diy 0=d\log(2\pi e)\left[\sum\limits_{\alpha =1,2}\lam_\alpha\BC_\alpha\bt
\exp\,\Big(\frac{1}{2}{\bt}^{\rm T}\BC_\alpha\bt \Big) -\BC\bt\exp\,\Big(\frac{1}{2}{\bt}^{\rm T}\BC\bt \Big)
\right]\\
\quad\diy +\sum\limits_{\alpha =1,2}\lam_\alpha
\log\, ({\rm{det}}\,\BC_\alpha)\BC_\alpha\bt\left[\exp\,\Big(\frac{1}{2}{\bt}^{\rm T}\BC_\alpha\bt \Big)\right]
-\log\,({\rm{det}}\,\BC)\BC\bt\left[\exp\,\Big(\frac{1}{2}{\bt}^{\rm T}\BC\bt \Big)\right]
\end{array}\eeq
or
\beq\beal\diy  \bt=\sum\limits_{\alpha =1,2}\lam_\alpha\BC^{-1}\BC_\alpha\bt
\left\{\exp\,\Big[\frac{1}{2}{\bt}^{\rm T}(\BC_\alpha -\BC)\bt \Big]\right\}\\
\qquad\qquad\qquad\qquad\qquad\qquad\diy\times\frac{d\log e+\log \left[ (2\pi)^d ({\rm{det}}\,\BC_\alpha)\right]
}{d\log\, e
}.\end{array}\eeq

An obvious solution is $\bt={\mathbf 0}$; we are tempting to suggest that it is unique. To analyze
the character of this point, let us take the second gradient:
\beq\label{eq:2ndgrad}\beal\diy\nabla^2_{\bt\bt}\Lam (\bt )=d\log(2\pi e)
\left[\sum\limits_{\alpha =1,2}\lam_\alpha(\BC_\alpha+ \BC_\alpha\bt\bt^{\rm T}\BC^{\rm T}_\alpha)
\exp\,\Big(\frac{1}{2}{\bt}^{\rm T}\BC_\alpha\bt \Big)\right.\\
\diy\qquad\qquad \qquad  -(\BC +\BC\bt\bt^{\rm T}\BC^{\rm T})\exp\,\Big(\frac{1}{2}{\bt}^{\rm T}\BC\bt \Big)
\Bigg]\\
\qquad\diy +\sum\limits_{\alpha =1,2}\lam_\alpha
\log\, ({\rm{det}}\,\BC_\alpha)(\BC_\alpha +\BC_\alpha\bt\bt^{\rm T}\BC_\alpha^{\rm T})\exp\,\Big(\frac{1}{2}{\bt}^{\rm T}\BC_\alpha\bt \Big)\\
\diy\qquad\qquad \qquad -\log\,({\rm{det}}\,\BC)(\BC +\BC\bt\bt^{\rm T}\BC^{\rm T})
\exp\,\Big(\frac{1}{2}{\bt}^{\rm T}\BC\bt \Big).\ena\eeq
At $\bt =0$ it yields
\beq\label{eq:2ndgrad0}\nabla^2_{\bt\bt}\Lam (\bt )\Big|_{\bt={\mathbf 0}}=\sum\limits_{\alpha =1,2}
\lam_\alpha
\log\, ({\rm{det}}\,\BC_\alpha)\BC_\alpha -\log\,({\rm{det}}\,\BC)\BC.\eeq
For a local minimum at the origin $\bt={\mathbf 0}$ we need the matrix $\sum\limits_{\alpha =1,2}
\lam_\alpha
\log\, ({\rm{det}}\,\BC_\alpha)\BC_\alpha -\log\,({\rm{det}}\,\BC)\BC$ to be positive definite. In other words,
the following property emerges, featuring reduced convexity
of the map $\BC\mapsto\BC\log\,{\rm{det}}\,\BC$: for given positive definite $\BC_1$, $\BC_2$,
 \def\bear{\begin{array}{r}}
\beq\label{eq:conve}\bear\lam\in [0,1]\mapsto [\lam\BC_1+(1-\lam )\BC_2]
\log\,{\rm{det}}\,[\lam\BC_1+(1-\lam )\BC_2]\qquad{}\\
\hbox{is a convex matrix-valued function.}\ena\eeq
Again, we can say that our numerical evidence suggests that $\bt =\b0$ can be a local extremum
(of either type) or a saddle point.
\vskip .5 truecm

{\bf 2.4.} Let us check the status of the origin in the case $d=1$. Here $\bt=t\in\bbR$, and $\BC_a =c_a$,
$a =1,2$, and $\BC=c=\lam_1c_1+\lam_2c_2$ are scalars. Further, function $\Lam (t)$ from
Eqn \eqref{eq:wKFL2} has the form
\beq\beal\Lam (t)= \diy \log(2\pi e)\left[\sum\limits_{\alpha =1,2}\lam_\alpha
\exp\,\Big(\frac{1}{2}c_\alpha t^2 \Big) -\exp\,\Big(\frac{1}{2}c t^2\Big)
\right]\\
\qquad\diy +\sum\limits_{\alpha =1,2}\lam_\alpha
(\log\,c_\alpha)\left[\exp\,\Big(\frac{1}{2}c_\alpha t^2 \Big)-1\right]
-(\log\,c)\left[\exp\,\Big(\frac{1}{2}ct^2 \Big)-1\right].\ena\eeq
Next, functions $F^{(1)}$ and $F^{(2)}$ from Eqn \eqref{eq:KyFanCond0} become
\beq\beal\diy F^{(1)}(t )=\sum\limits_{\alpha =1,2}\lam_\alpha\exp\,\Big(\frac{1}{2}c_\alpha t^2 \Big)
-\exp\,\Big(\frac{1}{2}c t^2 \Big),\\
\diy F^{(2)}(t )=\Bigg[\sum\limits_{\alpha =1,2}\lam_\alpha\exp\,\Big(\frac{1}{2}c_\alpha t^2 \Big)
-\exp\,\Big(\frac{1}{2}ct^2 \Big)\Bigg]\log\,(2\pi c)\\
\qquad\qquad\quad +\diy\sum\limits_{\alpha =1,2}\lam_\alpha\exp\,\Big(\frac{1}{2}c_\alpha t^2 \Big)
c^{-1}c_\alpha-\exp\,\Big(\frac{1}{2}ct^2 \Big).\ena\eeq
Correspondingly, for set $\bbS$ we obtain:
\beq\beal\bbS =\diy\Bigg\{t\in\bbR :\;\lam_1\exp\,\left[\frac{1}{2}\lam_2(c_1-c_2)t^2\right]
+\lam_2\exp\,\left[\frac{1}{2}\lam_1(c_2-c_1)t^2\right]\geq 1,\\
\quad\diy\bigg(\lam_1\exp\,\left[\frac{1}{2}\lam_2(c_1-c_2)t^2\right]
+\lam_2\exp\,\left[\frac{1}{2}\lam_1(c_2-c_1)t^2\right]-1\bigg) \log\,(2\pi c)\\
\diy\qquad\qquad +\diy\frac{\lam_1c_1}{c}\exp\,\left[\frac{1}{2}\lam_2(c_1-c_2)t^2 \right]
+\frac{\lam_2c_2}{c}\exp\,\left[\frac{1}{2}\lam_1(c_2-c_1)t^2 \right]\leq 1\Bigg\}.\ena\eeq
It is can be seen that if $0<c<1/(2\pi )$, set $\bbS$ is unbounded.

The stationary point is where $\diy\frac{\partial}{\partial t}\Lam =0$ or
$$\beal \diy 0=t\,\log(2\pi e)\left[\sum\limits_{\alpha =1,2}\lam_\alpha c_\alpha
\exp\,\Big(\frac{1}{2}c_\alpha t^2 \Big) -c\exp\,\Big(\frac{1}{2}c t^2\Big)
\right]\\
\quad\diy +t\sum\limits_{\alpha =1,2}\lam_\alpha c_\alpha
(\log\,c_\alpha)\exp\,\Big(\frac{1}{2}c_\alpha t^2 \Big)
-c(\log\,c)\exp\,\Big(\frac{1}{2}ct^2 \Big).\ena$$

Clearly, $t=0$ is a solution. Next, we calculate the second derivative
$\diy\frac{\partial^2}{\partial t^2}\Lam$ at $t=0$. A general expression is:
$$\beal \diy\frac{\partial^2}{\partial t^2}\Lam (t)=\log(2\pi e)\left[\sum\limits_{\alpha =1,2}\lam_\alpha c_\alpha (1+c_\alpha t^2)
\exp\,\Big(\frac{1}{2}c_\alpha t^2 \Big) -c(1+ct^2)\exp\,\Big(\frac{1}{2}c t^2\Big)
\right]\\
\quad\diy +\sum\limits_{\alpha =1,2}\lam_\alpha c_\alpha (1+c_\alpha t^2)
(\log\,c_\alpha)\exp\,\Big(\frac{1}{2}c_\alpha t^2 \Big)
-c(1+ct^2)(\log\,c)\exp\,\Big(\frac{1}{2}ct^2 \Big).\ena$$
Taking into account that $\diy\sum\limits_{\alpha =1,2}\lam_\alpha c_\alpha =c$, we obtain
that
\beq\label{eq:good}\left.\frac{\partial^2}{\partial t^2}\Lam (t)\right|_{t=0}=\sum\limits_{\alpha =1,2}\lam_\alpha c_\alpha
(\log\,c_\alpha)-c\log\,c\geq 0.\eeq
Inequality \eqref{eq:good} holds since $x\mapsto x\log\,x$ is a convex function for $x>0$.

\section{Numerical results}

Performing numerical simulations, we
show the graph of function $\bt\mapsto \Lam(\bt)$ in Eqn \eqref{eq:wKFL2}
for $d=2$ within chosen ranges of argument $\bt =\left(\beac t_1\\ t_2\ena\right)$ around
point $\bt ={\mathbf 0}$. Matrices
$\BC_a$ are taken in the form $\left(\diy\begin{array}{cc}\sigma_a^2&\rho_a\sigma_a^2\\
\rho_a\sigma_a^2&\sigma_a^2\end{array}\right)$ where $\sigma_a>0$ and $|\rho_a|<1$.

The simulations show the behavior of function $\Lam(\bt )$ for chosen matrices
$\BC_1$ and $\BC_2$
and values of $\lam_1 =\lam$ and $\lam_2=1-\lam$ over selected ranges of the argument
$\bt =\left(\beac t_1\\ t_2\ena\right)$ and indicate the set $\bbS$. The plots exhibit a
variety of possible patterns: positive and negative values of $\Lam(\bt )$, convexity,
concavity, global/local minimum/maximum, as well as a saddle point, at $\bt ={\mathbf 0}$.
(Recall, $\Lam>0$ has been proposed as an improvement whereas
$\Lam<0$ as a retrogression of a standard KFI.)

\begin{center}
\includegraphics[scale=1]{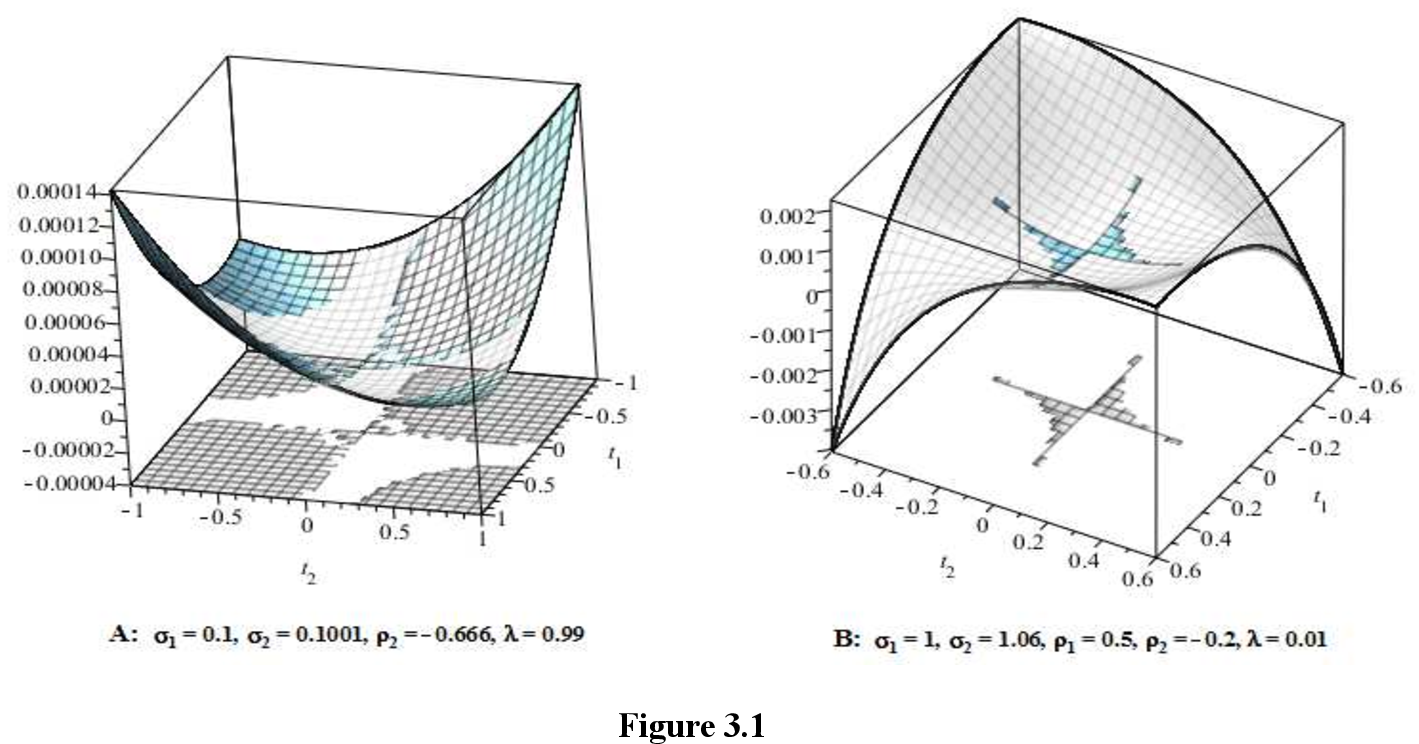}
\end{center}
\label{fig}

In Fgures 3.1 -- 3.5, set  $\bbS$ -- when it is non-empty --
is shown in a grey color at a bottom horizontal
plane. (The level at which
this plane is placed has been selected for presentational convenience only.) We want to note that in some examples the origin $\bt ={\mathbf 0}$ seems to be an isolated point  in $\bbS$:
it may be a consequence of the fact that the WF $\phi (\bx )=\exp\,\big(\bt^{\rm T}
\bx\big)$ is unbounded for $\bt\neq{\mathbf 0}$.

In Figures 3.1.A and 3.1.B the graph of
function $\Lam(\bt )$ lies above the value $0$ attained at $\bt ={\mathbf 0}$. This
suggests that $\bt ={\mathbf 0}$ is a global minimum of $\Lam(\bt )$. Consequently,
the matrix in the RHS of Eqn \eqref{eq:2ndgrad0} is positive definite for the specified choices of
$\BC_1$, $\BC_2$ and $\lam$.

Apparently, the EKFI holds true far beyond  $\bbS$ and yields an improvement of the standard KFI.
Matrix $\BC_1$ in Figure 3.1.A is of the form
$\BC_1=\left(\begin{array}{cc}\sigma_1^2&0\\
0&\sigma_1^2\end{array}\right)$ and $\BC_2=\left(\begin{array}{cc}\sigma_2^2&\sigma_2^2\rho_2\\
\sigma_2^2\rho_2&\sigma_2^2\end{array}\right)$. The values of $\lam=\lam_1$ are chosen
to be $0.99$ on Figure 3.1.A and $0.001$ on Figure 3.1.B.

\begin{center}
\includegraphics[scale=.8]{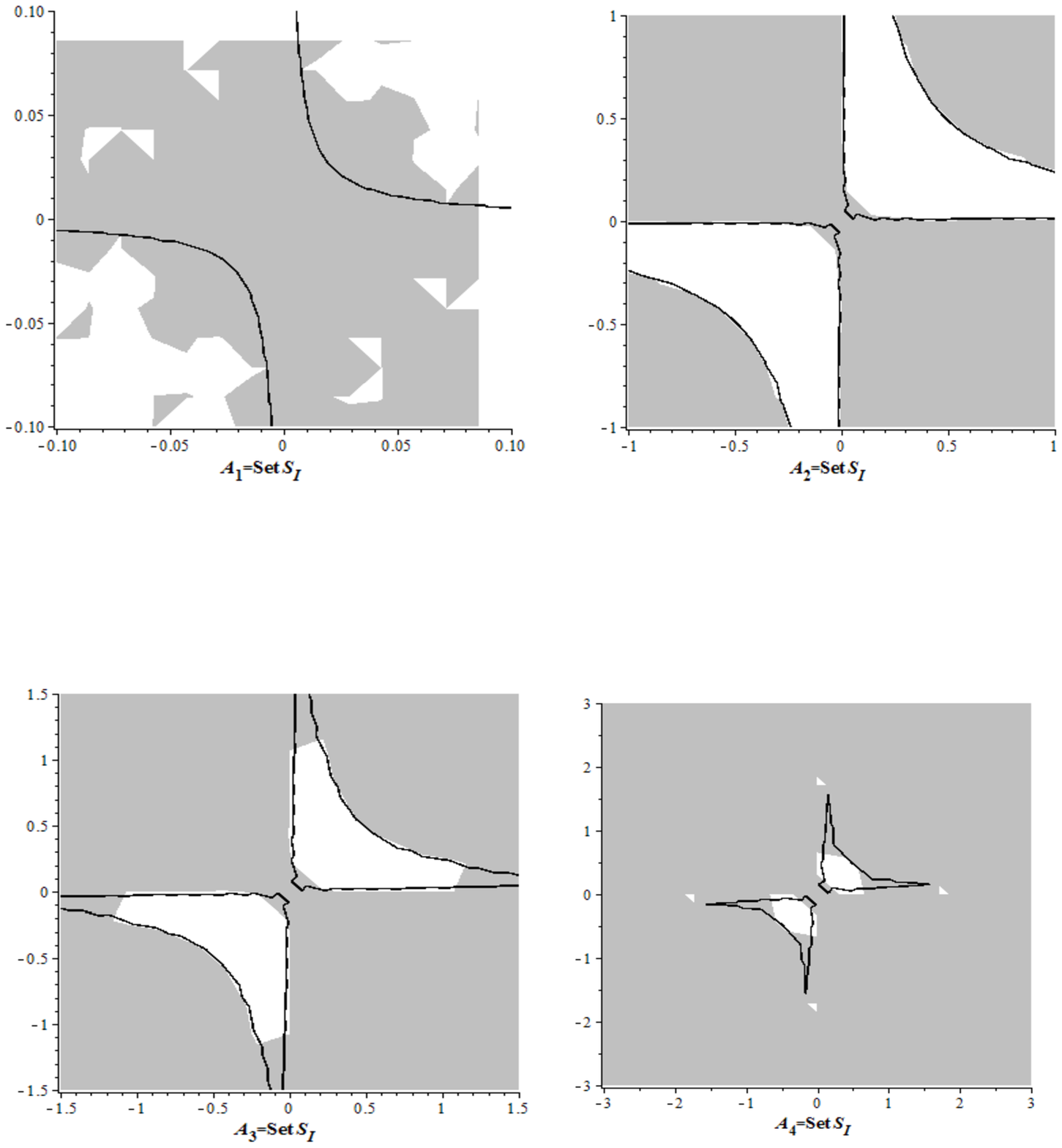}
\end{center}
\label{fig}

The plots above give an impression of set $\bbS$ in Figure 3.1.A showing how our perception
changes when the range of variables $t_1$ and $t_2$ increases. In this example, the set (shown in the
grey color) is, obviously, unbounded. (The hyperbola-type curves are used to provide a geometric
reference.)  On the other hand, the plots below demonstrate that set $\bbS$ in Figure 3.1.B
is bounded.

\begin{center}
\includegraphics[scale=.8]{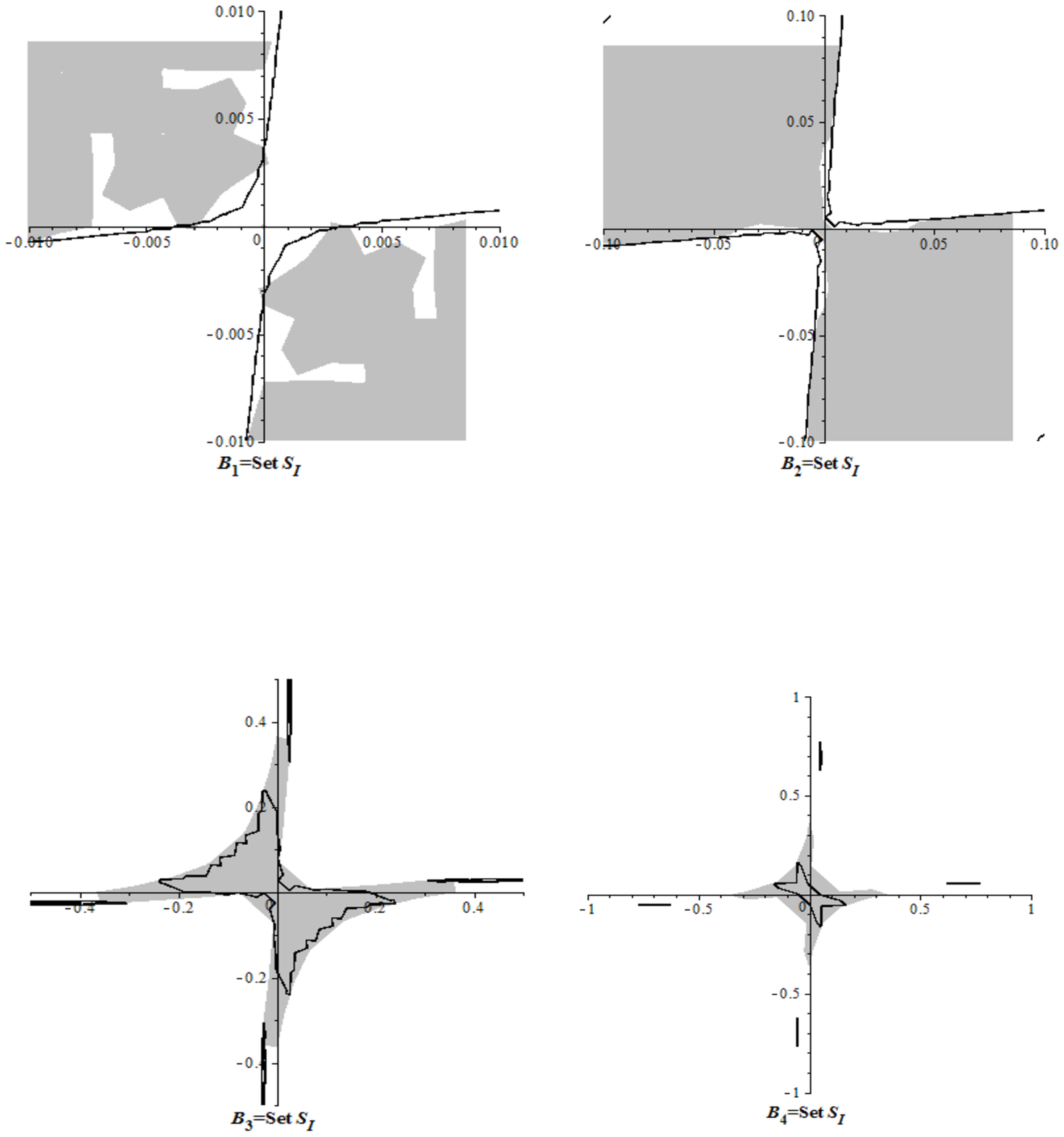}
\end{center}
\label{fig}

Next, Figure 3.2 shows a more complex character of behavior. Here $\bt ={\mathbf 0}$ is,
apparently, a saddle point for the graph of $\Lam(\bt )$. Function $\Lam(\bt )$
takes both positive and negative values. However, over set $\bbS$ the EKFI
yields an improvement of the standard KFI.
Here matrices $\BC_\alpha=\left(\begin{array}{cc}\sigma_\alpha^2&\sigma_\alpha^2\rho_\alpha\\
\sigma_\alpha^2\rho_\alpha&\sigma_\alpha^2\end{array}\right)$, $\alpha =1,2$, and
$\lam_1=\lam_2=\lam =1/2$.

Further, a similar pattern of behavior is confirmed on Figures 3.3.A and 3.3.B, with the same form
of matrices $\BC_\alpha$, and with $\lam_1=\lam =0.99$ and $0.5$, respectively.

\begin{center}
\includegraphics[scale=.7]{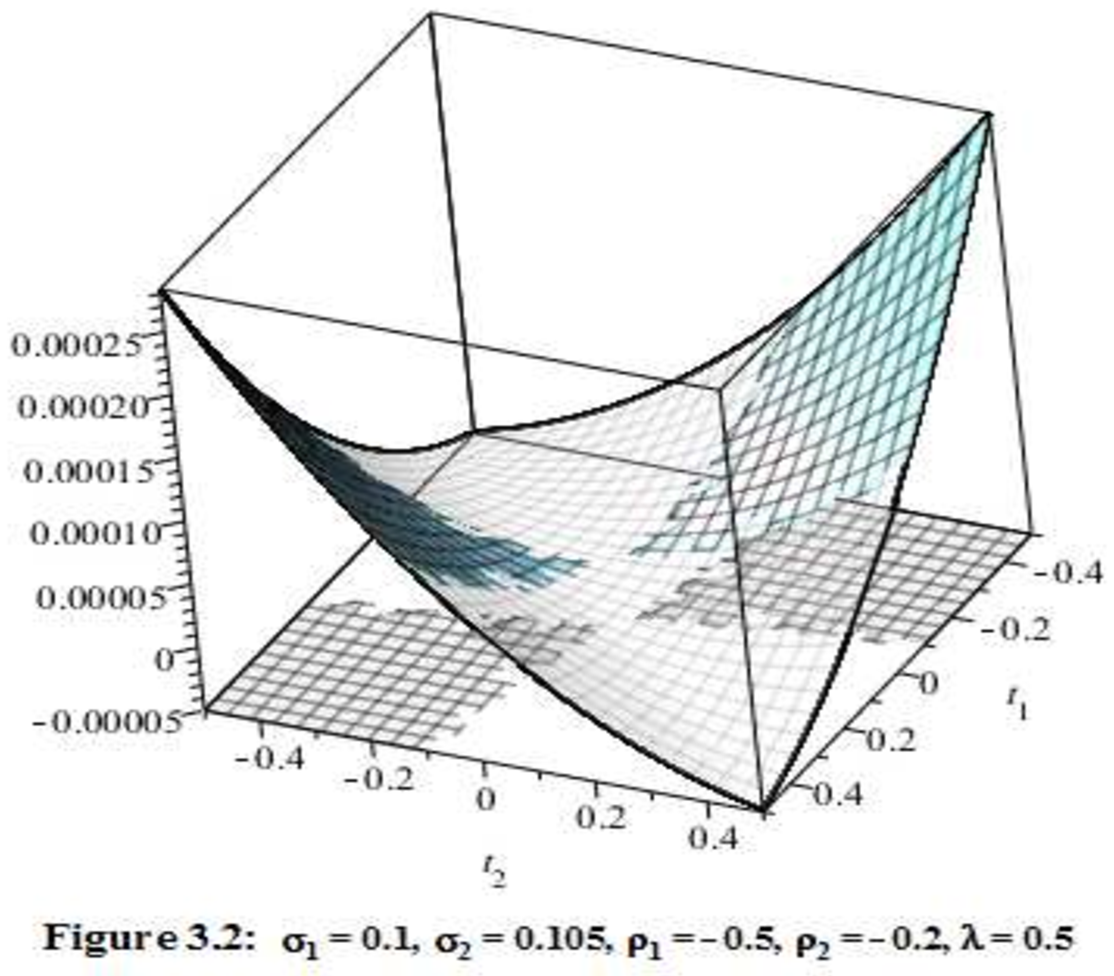}
\end{center}
\label{fig}

\begin{center}
\includegraphics[scale=1]{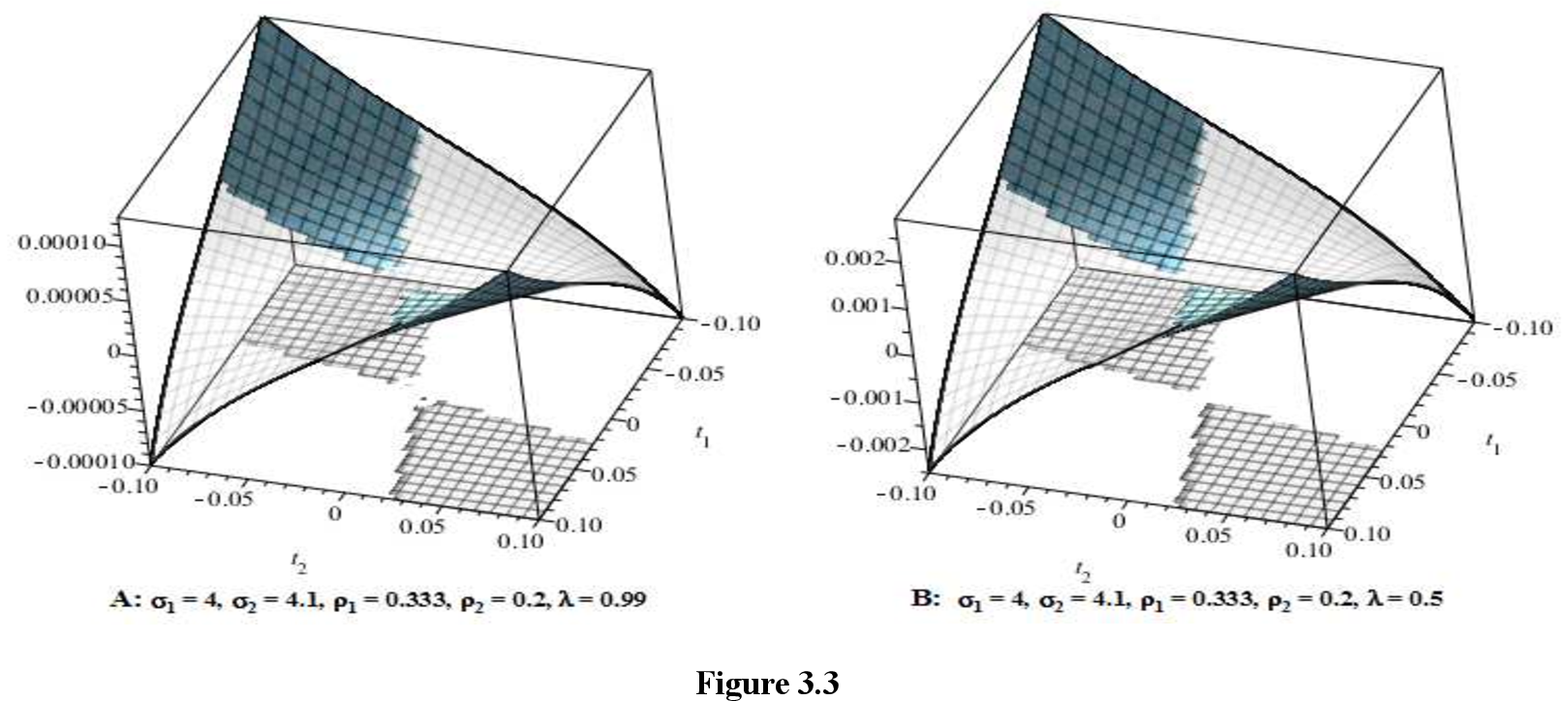}
\end{center}
\label{fig}

In Figure 3.4  we see an example where $\bbS$  is non-empty, and
$\Lam(\bt )<0$ for some $\bt\in\bbS$. In other words, this is an example where
the the EKFI holds true but does not yield an improvement relative
to the standard KFI. In this example, matrix $\BC_1 =\left(\begin{array}{cc}\sigma_1^2&0\\
0&\sigma_1^2\end{array}\right)$ and $\BC_2=\left(\begin{array}{cc}\sigma_2^2&\sigma_2^2\rho_2\\
\sigma_2^2\rho_2&\sigma_2^2\end{array}\right)$. The value $\lam=\lam_1=0.99$.

\begin{center}
\includegraphics[scale=.9]{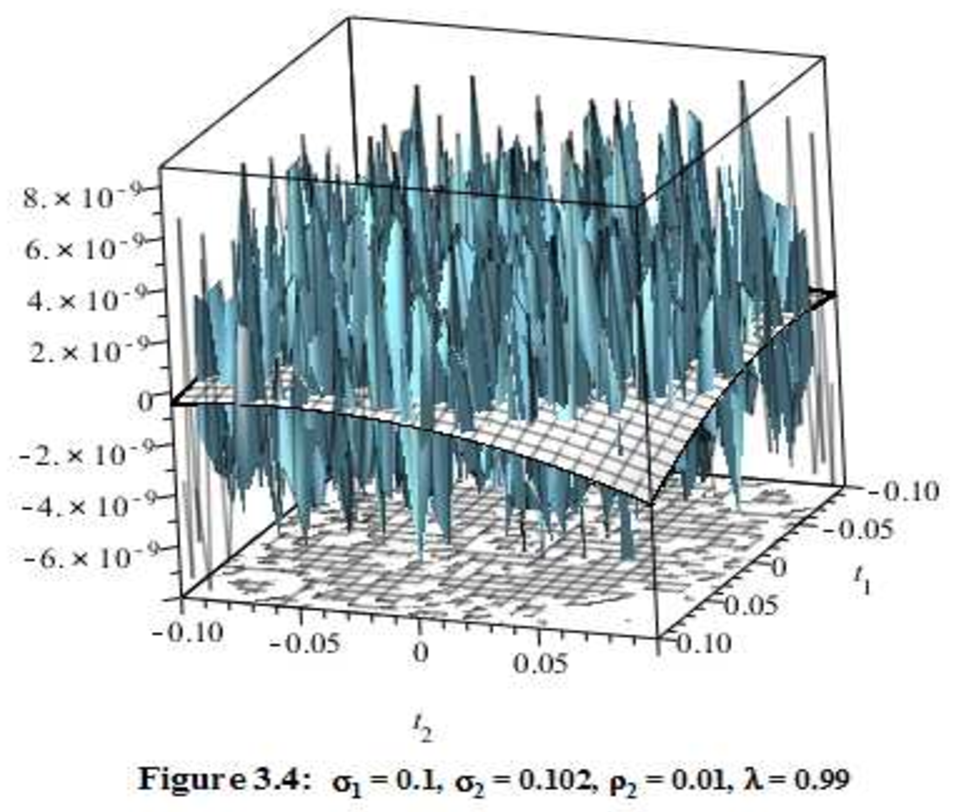}
\end{center}


\vskip .5 truecm

\begin{center}
{\includegraphics[scale=1]{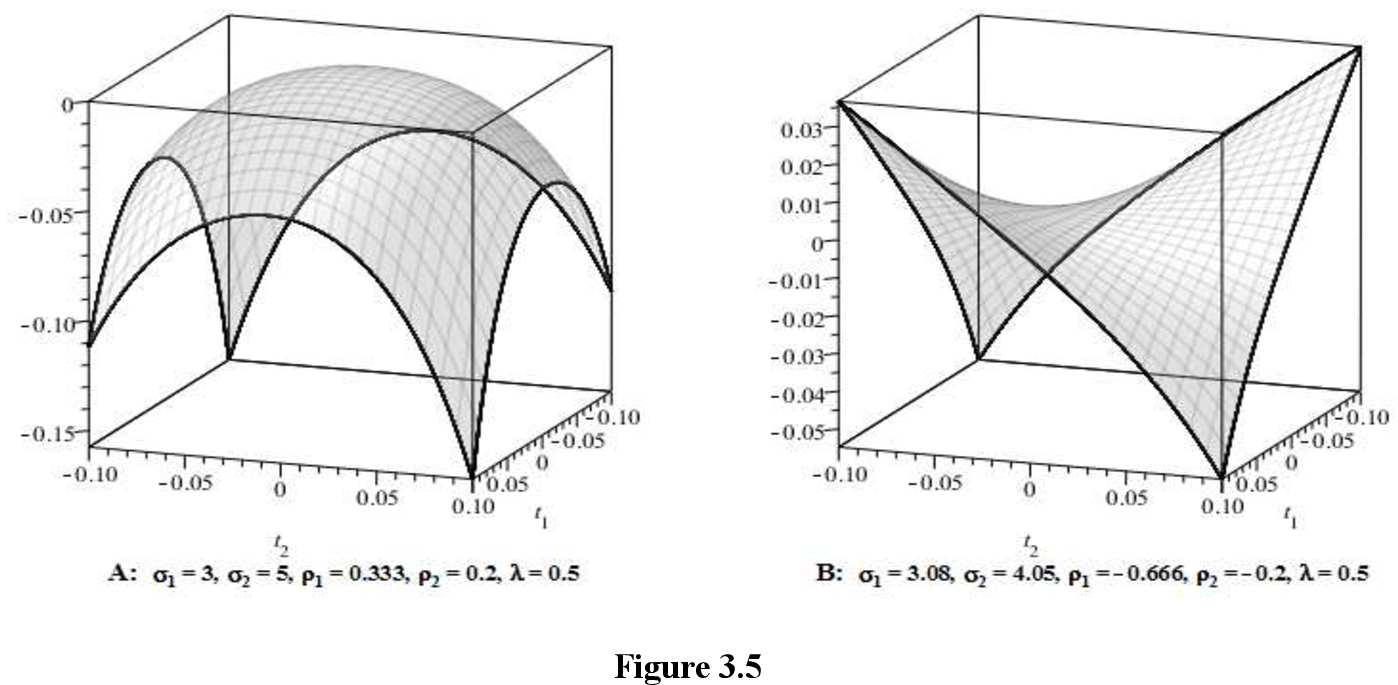}}
\end{center}
\label{fig}

Finally, in Figures 3.5 the set $\bbS$ is empty.  Consequently,
the EKFI fails (within the indicated range of argument
$\bt$). Also, function $\Lam$
in Figure 3.5.A takes negative values for $\bt\neq{\mathbf 0}$ within the indicated range but $\Lam$ in Figure 3.5.B takes both negative and positive values within the indicated range.
The matrices are $\BC_1=\left(\begin{array}{cc}\sigma_1^2&\sigma_1^2\rho_1\\
\sigma_1^2\rho_1&\sigma_1^2\end{array}\right)$ and $\BC_2=\left(\begin{array}{cc}
\sigma_2^2&\sigma_2^2\rho_2\\
\sigma_2^2\rho_2&\sigma_12^2\end{array}\right)$ and $\lam_1=\lam_2=\lam =1/2$. The
difference between these figures is that in Figure 3.5.A $\bt ={\mathbf 0}$ is a maximum of
$\Lam$ (within the depicted range of $\bt$) whereas in Figure 3.5.B the maximum
is at the corner points. 

\vskip .5 truecm
{\emph{Acknowledgements --}}
 YS thanks the Math Department, Penn State University,  for the financial support and hospitality
during the academic year 2014-5. SYS thanks the CAPES PNPD-UFSCAR Foundation
for the financial support in the year 2014-5. SYS thanks
the Federal University of Sao Carlos, Department of Statistics, for hospitality during the year 2014-5.


\end{document}